\documentclass[prb,twocolumn,showspaces,superscriptaddress]{revtex4}

\usepackage{natbib}
\usepackage{graphicx}
\usepackage{dcolumn}
\usepackage{amsmath}
\usepackage{color}

\begin{document}

\title{Magnetization and specific heat of the dimer system CuTe$_2$O$_5$}

\author {R.~M.~Eremina}
\author{T.~P.~Gavrilova}
\affiliation{E. K. Zavoisky Physical-Technical Institute, 420029
Kazan, Russia}
\author{A.~G\"{u}nther}
\author{Zhe Wang}
\affiliation{Experimental Physics V, Center for Electronic
Correlations and Magnetism, Institute for Physics, Augsburg
University, D-86135 Augsburg, Germany}
\author {R.~Lortz}
\affiliation{Department of Physics, The Hong Kong University of
Science \& Technology, Clear Water Bay, Kowloon, Hong Kong}

\author {M.~Johnsson}
\affiliation{Department of Materials and Environmental Chemistry,
Stockholm University, S-10691 Stockholm, Sweden}
\author{H.~Berger}
\affiliation{Institute de Physique de la Mati\`{e}re Complexe, EPFL,
CH-1015 Lausanne, Switzerland}
\author {H.-A. Krug von Nidda}
\author {J. Deisenhofer}
\author {A. Loidl}
\affiliation{Experimental Physics V, Center for Electronic
Correlations and Magnetism, Institute for Physics, Augsburg
University, D-86135 Augsburg, Germany}

%
\begin{abstract}
We report on magnetization and specific heat measurements on single-crystalline CuTe$_2$O$_5$. The experimental data are directly
compared to theoretical results for two different spin structures,
namely an alternating spin-chain and a two-dimensional (2D) coupled
dimer model, obtained by Das et al. [Phys. Rev. B \textbf{77}, 224437
(2008)]. While the analysis of the specific heat does not allow to
distinguish between the two models, the magnetization data is in
good agreement with the 2D coupled dimer model.
\end{abstract}

\maketitle

\section{Introduction}
\label{intro}
Quantum magnets with a dimerized singlet ground state are a
fascinating research field with phenomena ranging from spin-Peierls
transitions in antiferromagnetic spin $S=1/2$ chains like CuGeO$_3$ \cite{Hase} or TiOCl
\cite{Seidel03,Lemmens04,Zakharov06} to the Bose-Einstein condensation of magnons
in external magnetic fields as reported for the prototypical spin $S=1/2$ dimer systems BaCuSi$_2$O$_6$ \cite{Sebastian2006} and TlCuCl$_3$ \cite{Yamada2007} and also very recently for Sr$_3$Cr$_2$O$_8$ and Ba$_3$Cr$_2$O$_8$ \cite{Nakajima06,Kofu09a,Aczel09,Castro10}. The singlet ground state and the excited triplet state of the dimers are separated by an energy gap due to the different exchange energy between parallel and antiparallel aligned spins. This gap can be controlled by the external magnetic field, which splits the triplet due to the Zeeman effect in such a way that the lower triplet level meets the singulet giving rise to the observed Bose-Einstein condensation. \cite{Giamarchi2008}. This is a typical example of a quantum phase transition, i.e. it is driven by a non thermal control parameter, and, therefore, of high interest concerning the investigation of quantum critical behaviour.

In the majority of dimer systems the dominating intra-dimer superexchange
coupling is easily identified as the one mediated by the shared
ligands of the magnetic ions in the structural dimer. In some
systems, however, this nearest-neighbour superexchange coupling is very weak
and one has to consider more complex superexchange paths via several
neighboring ions. It has been shown that the resulting exchange
couplings can be of the same order of magnitude as the nearest-neighbour
superexchange \cite{Whangbo03}, with the spin-gapped CuTe$_2$O$_5$
-- the compound under investigation in the present work -- as one particular
example \cite{Hanke,Lemmens,Deisenhofer}. It is important to note that the influence of the
nonmagnetic lone-pair tetravalent chalcogenide ion (Te$^{4+}$) can significantly change the
superexchange interactions between the copper spins (Cu$^{2+}$: $S=1/2$) as seen for the case of related CuSe$_2$O$_5$ (with Se$^{4+}$) which reportedly forms a spin-chain like magnetic structure with an
antiferromagnetic transition at about 17~K \cite{Kahn,Janson09}.

In CuTe$_2$O$_5$, the situation turned out to be more complicated and the exchange geometry between the dimers (one- or two dimensional) remained unresolved so far \cite{Deisenhofer,Das}. Here we will show that our latest magnetization measurements are in favour of the two dimensional exchange geometry.

\section{Description of the Problem}

\begin{figure}[t]
\centering
\includegraphics[width=70mm,clip]{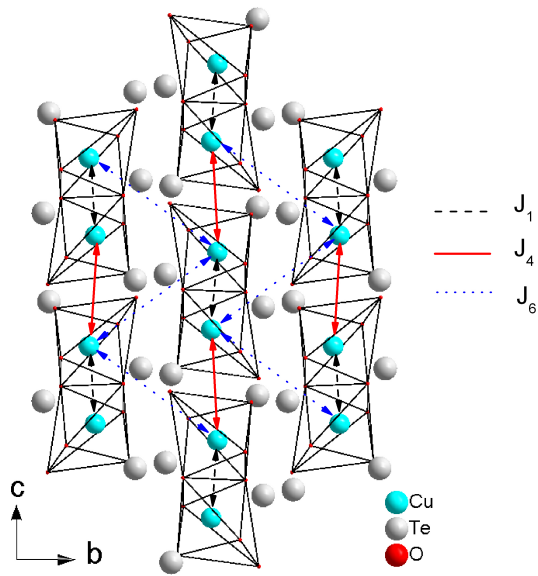}
\vspace{2mm}
\caption{\label{fig:structureFull}Projection of the monoclinic lattice structure
of CuTe$_2$O$_5$ with space group $P2_1$/c on the $bc$-plane. The
edge-sharing octahedra form Cu$_2$O$_{10}$ units which are separated
by Te ions. $J_1$, $J_4$, and $J_6$ indicate the most important exchange paths.}
\end{figure}

The crystalline structure of CuTe$_2$O$_5$ belongs to the space
group P2$_1$/c. The unit cell consists of four formula units, and
its dimensions are \emph{a} = 6.871 \AA, \emph{b} = 9.322 \AA,
\emph{c} = 7.602 \AA; the angle $\beta$ between the \emph{a} and
\emph{c} axes is 109.08$^\circ$ \cite{Hanke}. There are four
nonequivalent positions of copper ions in the unit cell. Each of the
copper ions is surrounded by six oxygen atoms forming a strongly
distorted octahedron. Neighboring Cu pairs form structural
Cu$_2$O$_{10}$ dimer units, built from edge sharing octahedra which
are rotated relative to each other and slightly magnetically nonequivalent.
The Cu$_2$O$_{10}$ units are connected via Te$^{4+}$ ions (see
Fig.~\ref{fig:structureFull}). Along the $b$ axis the Cu$_2$O$_{10}$ units exhibit alternating rotation with respect to each other resulting in strongly magnetic nonequivalent Cu sites in neighboring units.

The magnetic susceptibility of
CuTe$_2$O$_5$ exhibits a broad maximum at about 57~K and an exponential decrease to lower temperatures \cite{Lemmens,Deisenhofer}. These data have been successfully described in terms of the model of a
quasi-one-dimensional antiferromagnetic spin chain with alternating
exchange interactions, a modified Bleaney-Bowers
model \cite{Deisenhofer} or a two-dimensional (2D) system of coupled
dimers \cite{Das}. In all cases one obtains a leading exchange interaction of about 90~K,
followed by a second exchange contribution of the same order of
magnitude. The analysis of the susceptibility at the highest
temperatures yields a Curie-Weiss temperature $\Theta$ = -41~K \cite{Deisenhofer,Miljak}.
Electron spin resonance (ESR) studies
evidenced the existence of nonequivalent magnetic Cu sites from the
frequency dependence of the linewidth dur to the anisotropic Zeeman effect and, therefore, suggested that
strong superexchange bonds should exist only between equivalent
sites, while the superexchange between nonequivalent sites was
estimated to be of the order of
0.5~K \cite{Deisenhofer,Eremina,Gavrilova}.

Starting from the crystal structure the isotropic exchange
interactions of nine different pairs of copper ions have been
calculated by the Extended-H\"{u}ckel-Tight-Binding (EHTB) method
\cite{Deisenhofer}. With these calculations $J_6$ was identified as the largest exchange interaction followed by $J_1 = 0.59 J_6$ resulting in alternating spin chains. From the fitting of the susceptibility data this estimate was experimentally refined to $J_1 = 0.436 J_6$, with $J_6 = 93.3$~K. Later on the hopping integrals for the same pairs
of copper ions in CuTe$_2$O$_5$ were obtained using
density-functional calculations in combination with the $N$th order
muffin-tin orbital (NMTO) downfolding technique \cite{Das} and LDA+U
calculations \cite{Ushakov} and suggested that the system should be
regarded as a two-dimensional coupled dimer system with dominant
exchange coupling $J_4$ followed by $J_6 = 0.28 J_4$ and $J_1 = 0.11 J_4$.
Again these estimates were experimentally refined to $J_6 = 0.27 J_4$ and $J_1 = 0.07 J_4$ with $J_4=94.2$~K.
Das and coworkers \cite{Das} proposed to distinguish between the two magnetic
models by their contribution to the magnetization and the specific
heat. In this work we compare both quantities the magnetization and
the specific heat to these theoretical predictions.
Note that both models postulate a strong exchange coupling $J_6$ between significantly nonequivalent copper sites, which seems to be in contradiction to the ESR results. We will comment on this problem after evaluation of the data.

\section{Experimental Details}
\label{Experiment}
Single crystals of CuTe$_2$O$_5$ have been grown by a standard chemical vapor phase method. Mixtures of high purity CuO (Alfa-Aesar, 99.995\%) and TeO$_2$ (Acros, 99.9995\%) powders with a molar ratio higher than 1 : 2 were sealed in quartz tubes with HBr as transport gas for the crystal growth. Then the ampoules were placed horizontally into a tubular two-zone furnace and heated very slowly by 50$^\circ$C/h to 480$^\circ$C for about 3 days under high vacuum. The optimum temperatures at the source and deposition zones for the growth of single crystals have been 580$^\circ$C and 450$^\circ$C, respectively. After two months, large blue-green plates with a maximum size of $8 \times 8 \times 0.5$~mm were obtained. X-ray powder diffraction of these crystals revealed the proper single-phase product.

Magnetization measurements were performed in a DC--susceptometer
(Oxford Instruments, Teslatron) in fields up to 14~T and in the
temperature range 3-280~K. The temperature dependence of the
specific heat of CuTe$_2$O$_5$ was measured in a Physical Property
Measurement System PPMS (Quantum Design) between 1.8 K and 300~K.

\section{Results and Discussion}

\subsection{Magnetization}

\begin{figure}[b]
\centering
\includegraphics[width=70mm,clip]{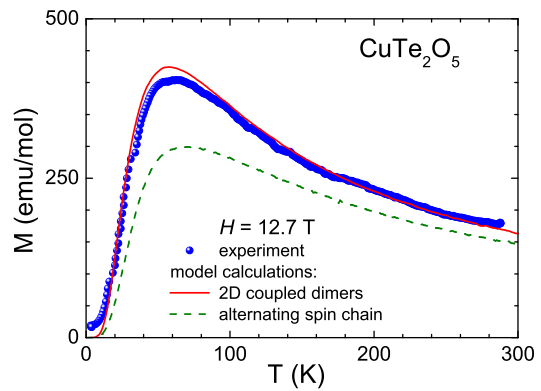}
\vspace{2mm} \caption{\label{fig:MvsT}Temperature dependence of the magnetization
measured in 12.7~T of CuTe$_2$O$_5$. The solid and dashed lines are
calculations based on an alternating spin chain ($J_6=93.3$~K) and a 2D coupled
dimer model ($J_4=94.2$~K) taken from Das {\it et al.} using the experimentally refined ratios of the exchange constants \cite{Das}.}
\end{figure}

The predictions given by Das \textit{et al.} \cite{Das} for the magnetization of the 2D coupled dimer model were
calculated for applied magnetic fields with strengths $h/J$ = 0.2, 0.5,
1.0 in units of the dominant exchange coupling $J$ which correspond
to absolute magnetic field values $H$ = 12.7, 31.7, 63.4~T, respectively.
Correspondingly, the magnetization for the alternating chain model
has been calculated for $H$ = 12.7 and 31.7~T for a direct
comparison \cite{Das}. Qualitatively both models yield a similar temperature dependence: starting from zero at $T=0$ due to the non magnetic singlet ground state of the spin dimers, the magnetization strongly increases with increasing temperature, when the dimers are breaking up, develops a maximum around the temperature corresponding to the antiferromagnetic exchange coupling $T_{\rm max} \sim J/k_{\rm B}$, and finally decreases following a Curie-Weiss law. However, quantitatively the absolute value of the maximum of the alternating spin-chain model (which is slightly shifted to higher temperatures) is about 25\% below that of the 2D coupled dimer model.

Only the lowest magnetic field is located in the range
reachable in our standard experimental laboratory setup, while the
higher fields have to be verified in specialized high-field
facilities. In Fig.~\ref{fig:MvsT} we show the data obtained at
12.7~T together with the predicted curves for the two models. The
agreement of the experiment with the prediction for the 2D coupled
dimer system is obviously much better than with the one for the alternating
spin chain based on ETHB calculations \cite{Deisenhofer}.
Therefore, the magnetization data clearly favour the 2D coupled
dimer model as a realistic description of the magnetic structure of
CuTe$_2$O$_5$.

\subsection{Specific Heat}

\begin{figure}
\centering
\includegraphics[width=70mm,clip]{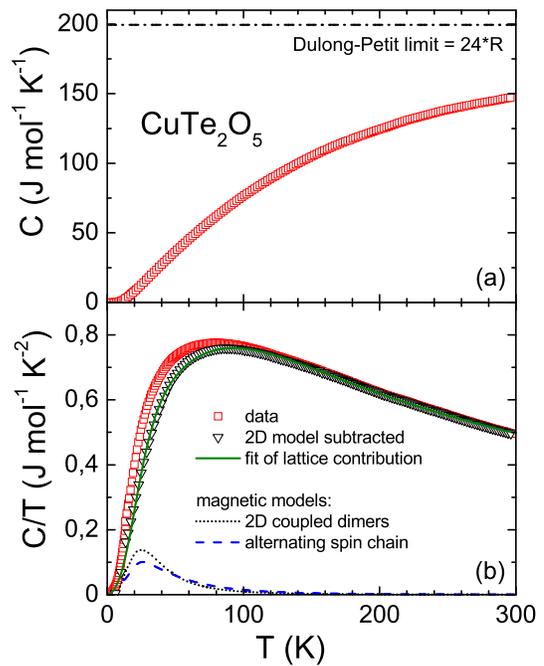}
\vspace{2mm} \caption{\label{Fig:Cdata} Temperature dependence of (a) the specific
heat and (b) the specific heat divided by temperature (squares) together with
theoretically suggested magnetic contributions for an alternating
spin chain $C_{alt}/T$ (dashed) and a system of 2D coupled dimers $C_{2D}/T$ (dotted) in CuTe$_2$O$_5$. The
dash-dotted line in (a) indicates the high-temperature Dulong-Petit
limit. The triangles in (b) denote the data after subtraction of the magnetic 2D coupled dimer contribution fitted by a pure lattice contribution (solid line).}
\end{figure}

In Fig.~\ref{Fig:Cdata}(a) we show the temperature dependence of the
specific heat. No anomalies corresponding to phase transitions have
been detected in agreement with reported susceptibility and ESR
measurements \cite{Deisenhofer}. As a result the specific heat
increases monotonously with increasing temperature. Notably, at
300~K the specific heat is still considerably lower than the
expected high-temperature value $3Rs = 200$~J/(mol K) for the phonon
contribution given by the Dulong-–Petit law indicating contributions
to the phonon-density of states from higher-lying lattice modes.
Here, $R$ denotes the gas constant and $s$ the number of atoms per
formula unit.

We assume that the total heat capacity originates from two different
contributions, a lattice contribution $C_{latt}$ due to acoustic and
optical phonons and  a magnetic contribution $C_{mag}$ corresponding
to the thermal population of excited dimer states. The magnetic
contributions to the specific heat divided by temperature for a
spin chain model $C_{alt}/T$ based on ETHB calculations \cite{Deisenhofer}
and for a 2D coupled dimer system
$C_{2D}/T$, as predicted by Das and coworkers, are plotted in
Fig.~\ref{Fig:Cdata}(b) in comparison to the total specific heat
$C/T$. Evidently, both magnetic contributions are small compared to
the lattice contribution and a non-magnetic reference material is
not available. Therefore, a straightforward method to unambiguously
extract the magnetic contribution from the experimental data is
difficult to realize.

Thus, we decided to chose the reverse approach by
subtracting the theoretically predicted magnetic contribution for
the 2D coupled dimer model, which is favoured by the magnetization measurements,
from the total specific heat and to analyze the residual
lattice contribution $C_{latt}$. The resulting data were approximated following standard procedures \cite{Gopal1966} with a minimized set of fit parameters only using a sum of one isotropic Debye (D) -- accounting for the 3 acoustic phonon branches -- and four isotropic Einstein terms ($E_{1,2,3,4}$) -- averaging the $3s-3=21$ optical phonon branches (fitting with less than four Einstein terms was not sufficient). For further reducing the number of free fit parameters, the ratio between these terms was fixed to $D:E_1:E_2 :E_3:E_4=1:1:2:2:2$ to account for the $3s=24$ degrees of freedom per formula unit. For $s=8$ atoms formula unit, the ratio between acoustical (Debye) and optical (Einstein) contributions is naturally fixed as 1:7. The weight distribution between the optical contributions is chosen in such a way that the lowest Einstein mode is of equal weight with the Debye contribution, assuming one low-lying isotropic optical phonon branch (with 3 degrees of freedom), while the remaining degrees of freedom have been equally distributed between the higher Einstein modes. The resulting fit curve (solid line in Fig.~\ref{Fig:Cdata}b) describes the data satisfactorily. For the respective Debye and Einstein temperatures we obtained $\Theta_{D}=144$~K, $\Theta_{E1}=151$~K, $\Theta_{E2}=308$~K, $\Theta_{E3}=497$~K, and $\Theta_{E4}=1225$~K.

To check the fit results of the lattice contribution we compare the optical phonon frequencies detected by Raman spectroscopy in CuTe$_2$O$_5$ in the infrared region between $100 \leq \nu \leq 900$~cm$^{-1}$ to the Einstein temperatures obtained from the specific heat. The leading Raman peaks and corresponding characteristic temperatures (given in brackets) are found at $\nu = 121$~cm$^{-1}$ (174~K), 211~cm$^{-1}$ (304~K), 375~cm$^{-1}$ (540~K), 444~cm$^{-1}$ (640~K), 745.5~cm$^{-1}$ (1074~K) \cite{Agazzi2007}. As one can see, the Einstein temperatures are in reasonable agreement with our fitting results, especially the low-lying optical mode at 174~K corresponds to $\Theta_{E1}=151$~K and also the existence of high-frequency modes at 640~K and 1074~K agrees well with the fact that the Dulong-Petit value is approached only far above room temperature.

\section{Discussion}

So far the present magnetization and specific-heat investigations favour the 2D coupled dimer model for CuTe$_2$O$_5$. The only open question remains concerning the frequency dependence of the ESR linewidth which was explained in terms of an anisotropic Zeeman effect requiring a very small exchange coupling between neighboring nonequivalent copper sites, whereas both theoretical models derive a rather strong exchange between these sites. This dilemma can be probably resolved regarding very recent experimental findings.

Current terahertz spectroscopic investigations of the singlet-triplet excitations in the dimerised low-temperature phase proof the importance of the Dzyaloshinsky-Moriya (DM) interaction in CuTe$_2$O$_5$ \cite{Wang2011} which has not been taken into account in the previous analysis of the ESR linewidth. Moreover, very recently the anisotropy as well as the temperature and frequency dependence of the ESR linewidth in the related uniform spin-1/2-chain compound CuSe$_2$O$_5$ have been successfully explained to result from the DM interaction \cite{Herak2011}. Due to the theory of Oshikawa and Affleck, which models the spin relaxation in uniform spin-1/2-chains at low temperature, the linewidth should increase with decreasing temperature and increasing frequency (i.e. resonance field) in the presence of a staggered field arising from the DM interaction \cite{Oshikawa1999,Oshikawa2002}. It is especially important to note that the DM contribution to the linewidth increases proportional to the square of the applied frequency and corresponding resonance field like the contribution of the anisotropic Zeeman effect. Thus, both contributions are difficult to separate.

Looking at the cases of copper benzoate \cite{Okuda1972}, where Oshikawa and Affleck first applied their theory, and also CuSe$_2$O$_5$ \cite{Herak2011}, it turns out that in both systems strong broadening with increasing frequency appears for the external field along a certain crystal axis, while the broadening is significantly weaker for any perpendicular direction. Despite the fact that CuTe$_2$O$_5$ is not a uniform spin-1/2-chain, but a complex dimer system, the pronounced line broadening with increasing frequency for the magnetic field applied along the $b$ axis compared to minor effects for the perpendicular orientations, suggests an analogous importance of the DM interaction for the spin relaxation in this compound. For a deeper analysis additional theoretical effort is needed to check the applicability of the Oshikawa-Affleck theory to the case of CuTe$_2$O$_5$. At the moment we can state that the anisotropic Zeeman effect alone is probably not enough to explain the frequency dependence of the linewidth and, hence, the exchange coupling between nonequivalent copper sites is not necessarily small. Insofar the seeming contradiction of the ESR data to the existing models is not mandatory. This finding supports our result that CuTe$_2$O$_5$ is best described in terms of a 2D coupled dimer model.

\section{Conclusion}

To summarize, the temperature dependent magnetization data of CuTe$_2$O$_5$ taken in an external magnetic field of 12.7~T agree well with the prediction for the 2D-coupled dimer model and clearly deviate from the expectation for the alternating spin-chain model. After subtraction of the corresponding magnetic specific-heat contribution from the experimental heat-capacity data the residual specific heat can be satisfactorily fitted by phonon-contributions only, which are in line with the characteristic phonon energies obtained from Raman spectroscopy. Thus, the two-dimensional model is clearly favoured by the present experiments.

\begin{acknowledgments}
We thank Thomas Wiedenmann for experimental support and Andrei Pimenov  (TU Vienna) for useful discussions. This work was supported by the Deutsche Forschungsgemeinschaft (DFG) via the Transregional Collaborative Research Center TRR 80 (Augsburg, Munich) and by the Russian Federal Program  N 02.740.11.0103.
\end{acknowledgments}


\begin{thebibliography}{29}

\bibitem{Hase}
M. Hase, I. Terasaki, and K. Uchinokura, Phys. Rev. Lett. \textbf{70},
3651 (1993).

\bibitem{Seidel03} A. Seidel, C. A. Marianetti, F. C. Chou, G. Ceder, and P. A. Lee,
Phys. Rev. B \textbf{67}, 020405 (2003).

\bibitem{Lemmens04} P. Lemmens, K. Y. Choi, G. Caimi, L. Degiorgi, N. N. Kovaleva, A.
Seidel, and F. C. Chou, Phys. Rev. B \textbf{70}, 134429 (2004).

\bibitem{Zakharov06} D. V. Zakharov, J. Deisenhofer, H.-A. Krug von Nidda, P.
Lunkenheimer, J. Hemberger, A. Loidl, R. Claessen, M. Hoinkis, M.
Klemm, M. Sing, M. V. Eremin, S. Horn, and A. Loidl, Physical Review
B \textbf{73}, 094452 (2006).

\bibitem{Sebastian2006} S. E. Sebastian, N. Harrison, C. D. Batista, L. Balicas, M. Jaime, P. A. Sharma, N. Kawashima, and I. R. Fisher, Nature \textbf{441}, 617 (2006).

\bibitem{Yamada2007} F. Yamada, T. Ono, H. Tanaka, G. Misguich, M. Oshikawa, and T. Sakakibara, J. Phys. Soc. Jpn. \textbf{77}, 013701 (2007).

\bibitem{Nakajima06} T. Nakajima, H. Mitamura, and Y. Ueda, J. Phys. Soc. Jpn. \textbf{75}, 054706 (2006).

\bibitem{Kofu09a} M. Kofu, J. H. Kim, S. Ji, S. H. Lee, H. Ueda, Y. Qiu, H. J. Kang, M. A. Green, and Y. Ueda, Phys. Rev. Lett. \textbf{102}, 037206 (2009).

\bibitem{Aczel09} A. A. Aczel, Y. Kohama, C. Marcenat, F. Weickert, M. Jaime,
O. E. Ayala-Valenzuela, R. D. McDonald, S. D. Selesnic, H. A. Dabkowska, and G. M. Luke, Phys. Rev. Lett. \textbf{103}, 207203 (2009).

\bibitem{Castro10} D. L. Quintero-Castro, B. Lake, E. M. Wheeler, A. T. M. N. Islam,
T. Guidi, K. C. Rule, Z. Izaola, M. Russina, K. Kiefer, and Y. Skourski, Phys. Rev. B \textbf{81}, 014415 (2010).

\bibitem{Giamarchi2008} T. Giamarchi, C. Rüegg, and O. Tchernyshyov, Nature Physics \textbf{4}, 198 (2008).


\bibitem{Whangbo03} M.-H. Whangbo, H.-J. Koo, D. Dai, and D. Jung, Inorg. Chem. \textbf{42}, 3898
(2003).

\bibitem{Hanke}
K. Hanke, V. Kupcik, and O. Lindqvist, Acta Crystal- logr., Sect.
B: Struct. Crystallogr. Cryst. Chem. \textbf{29}, 963 (1973).

\bibitem{Lemmens}
P. Lemmens, G. G\"{u}ntherodt, and C. Gros, Phys. Rep. \textbf{375}, 1
(2003).


\bibitem{Deisenhofer}
J. Deisenhofer, R. M. Eremina, A. Pimenov, T. Gavrilova, H. Berger,
M. Johnsson, P. Lemmens, H.-A. Krug von Nidda, A. Loidl, K.-S. Lee,
and M.-H. Whangbo, Phys. Rev. B: Condens. Matter \textbf{74}, 174421
(2006).


\bibitem{Kahn} O. Kahn, M. Verdaguer,  J. J. Girerd, J. Galy, and F.
Maury, Solid State Commun. \textbf{34}, 971 (1980).


\bibitem{Janson09} O. Janson, W. Schnelle, M. Schmidt, Yu. Prots, S.-L. Drechsler, S.
K. Filatov, H. Rosner, New J. Phys. \textbf{11}, 113034 (2009).

\bibitem{Das}
H. Das, T. Saha-Dasgupta, C. Gros, and R. Valenti, Phys. Rev. B \textbf{77}, 224437 (2008).


\bibitem{Miljak}
M. Miljak, M. Herak, O. Milat, N. Tomasic, and H. Berger, J. Phys.:
Condens. Matter \textbf{20}, 505210 (2008).

\bibitem{Eremina}
R. M. Eremina, T. P. Gavrilova, H.-A. Krug von Nidda, A. Pimenov, J.
Deisenhofer, and A. Loidl, Physics of the Solid State \textbf{50}, 283(2008).

\bibitem{Gavrilova}
T. P. Gavrilova, R. M. Eremina, H.-A. Krug von Nidda, J.
Deisenhofer, A. Loidl, Journal of optoelectronics and advanced
materials \textbf{10}, 1655 (2008).

\bibitem{Ushakov} A. V. Ushakov and S. V. Streltsov, J. Phys.: Condens. Matter
\textbf{21}, 305501 (2009).

\bibitem{Gopal1966} E. S. R. Gopal: \textit{Specific heats at low temperatures}, Heywood, London (1966).

\bibitem{Agazzi2007} L. Agazzi, Master Thesis, Pavia (2007).

\bibitem{Wang2011} Z. Wang, M. Schmidt, Y. Goncharov, Y. Skourski, J. Wosnitza, H. Berger, H.-A. Krug von Nidda, A. Loidl, and J. Deisenhofer, J. Phys. Soc. Jpn. \textbf{80}, 124707 (2011).

\bibitem{Herak2011} M. Herak, A. Zorko, D. Arcon, A. Potocnik, J. van Tol, A. Ozarowski, and H. Berger, arXiv:11095597 (2011).

\bibitem{Oshikawa1999} M. Oshikawa and I. Affleck, Phys. Rev. Lett. \textbf{82}, 5136 (1999).

\bibitem{Oshikawa2002} M. Oshikawa and I. Affleck, Phys. Rev. B \textbf{65}, 134410 (2002).

\bibitem{Okuda1972} K. Okuda, H. Hata, and M. Date, J. Phys. Soc. Jpn. \textbf{33}, 1574 (1972).



\end{thebibliography}
\end{document}